\documentclass[showpacs,prl,aps,twocolumn]{revtex4}
\begin{document}
\title{Evidence of ratchet effect in nanowires of a conducting polymer}
\author{A. Rahman}
\author{M. K. Sanyal}
\author{R. Gangopadhayy}
\author{A. De}
\author{I. Das}
\affiliation{{\it Saha Institute of Nuclear Physics, 1/AF
Bidhannagar, Kolkata 700 064, India.}}
\date{\today}
\begin{abstract}
Ratchet effect, observed in many systems starting from living
organism to artificially designed device, is a manifestation of
motion in asymmetric potential. Here we report results of a
conductivity study of Polypyrrole nanowires, which have been
prepared by a simple method to generate a variation of doping
concentration along the length. This variation gives rise to an
asymmetric potential profile that hinders the symmetry of the
hopping process of charges and hence the value of measured
resistance of these nanowires become sensitive to the direction of
current flow. The asymmetry in resistance was found to increase
with decreasing nanowire diameter and increasing temperature. The
observed phenomena could be explained with the assumption that the
spatial extension of localized state involved in hopping process
reduces as the doping concentration reduces along the length of
the nanowires.
\end{abstract}
\maketitle
Flow of particles becomes direction sensitive in presence of a
ratchet potential whose principal feature is loss of inversion
symmetry \cite {1,2,3,4,5,6,7,8,9,10,11}. Depending upon the
origin of asymmetric potential and fluctuating force several kinds
of ratchets are observed ranging from molecular motors \cite {3}
in which proteins move in a deterministic way along filaments, to
electron pumps \cite {5} engineered in semiconductor channels.
Ratchet effect has been utilized in diverse fields like in the
process of particle separation \cite {8} and in optical tweezing
\cite {9}. We report here an evidence of ratchet potential
formation in nanowires of a conducting polymer. The resistance of
these nanowires can differ by several kilo-Ohms as the direction
of current flow along the length is reversed. This effect become
more pronounced with decrease of diameter of the nanowires and
increase in temperature.

Strong dependence of conductivity ($\sigma$) of conjugated
polymers on doping concentration ($c$), defined as number of
carriers per monomer, is indeed an important phenomenon to be
exploited in polymer electronics. Experimentally observed seven
orders of magnitude increase in $\sigma$ due to increase of $c$
from $0.005$ to $0.2$ has been explained with a variable range
hopping (VRH) theory by invoking doping concentration dependent
size of the extent of localized region that vary from $1$ to $5$
nm \cite {12}. We have used a simple preparation process in which
the gradient of $c$ along polymer wires could be controlled by
performing the polymerization reaction in a confined environment,
provided by pores of membranes. Polycarbonate membranes (Whatman
Inc.) of thickness $\sim 6 \mu m$ was placed between two
compartments of a chemical cell having aqueous solution of pyrrole
monomer ($0.1$M) in one side and ferric chloride ($FeCl_3$) (0.5M)
as oxidizing agent in the other compartment. The oxidizing agent
$FeCl_3$ acts as polymerization agent in formation of Polypyrrole
and provides dopant (counter anion) $Cl^-$. The atomic ratio of
$Cl$ to $N$ determines the degree of dopant and one obtains
$c=0.33$ for fully doped Polypyrrole \cite{doping}.

Several membranes each having uniform pore diameter ranging from
30 nm to 200 nm were used for growing nanowires of different
diameters. Each membrane was first exposed to monomer to fill the
pores and then after about five minutes $FeCl_3$ solution was
allowed to flow in the pores. Polymerization raction takes place
within each pore as $FeCl_3$ start diffusing through these pores
towards the compartment containing the monomer. This diffusion
process of the oxidizing agent create a profile of $c$ that
reduces continuously along the length of the nanowire as shown in
schematic diagram Fig. 1(a). The results of Scanning Electron
Microscopy (SEM) and Transmission Electron Microscopy (TEM)
measurements performed at various stages of this growth process
indicated that polypyrrole preferentially nucleate to the pore
wall forming tubes first and then the tubes get converted into
solid wire (Fig. 1(b)) as observed earlier \cite {13}.We have
extracted dopant ($Cl^-$) concentration as a function of depth
using Secondary Ion Mass Spectroscopy (SIMS) technique. SIMS
profiles were extracted for $Cl$ and $N$ from both faces of
membranes to confirm consistency of the profiles. More than two
orders of magnitude variation of $c$  was observed along the
length of nanowires having 30 nm diameter and constant doping was
obtained only at a depth of $400$ nm from $FeCl_3$ compartment.
But for $200$ nm diameter nanowires the variation was found to be
only one order of magnitude and constant doping was obtained
within $100$ nm depth (Fig.1(c)).

To determine the current-voltage (I-V) characteristics of
nanowires a direct current was swept between positive and negative
maximum values across the membrane by applying silver paste on
both sides. The nanowires in the membrane get connected in
parallel and the voltage developed across the nanowires was
measured in this pseudo four-probe geometry (Fig. 2(a) (inset)).
In Fig. 2(a) we have shown typical resistance data of a membrane
having 30nm nanowires where direction of current ($100\mu$ Amp for
297.5K, 240K and $1\mu$ Amp for 133K and 64K) is reversed at
$t=50$ sec. Significant difference ($\sim 67$ kilo-ohms) in
measured value of resistance was observed even at 64K data though
the effect was found to be more at higher temperatures. In Fig.
2(b) we have presented I-V data measured with positive ($I^+$),
and negative ($I^-$), current to illustrate the effect of
temperature. We have repeated the measurement with conducting
copper tape and gold sputtered contact and obtained similar
results. This observation and the fact that observed resistance
asymmetry reduces with lowering of temperature rules out the
possibility that asymmetry in resistance is arising due to the
formation of Sch\"ottky barrier in the contact. We observed that
best quality samples could be prepared by cooling the membrane at
liquid nitrogen temperature right after the nanowire fabrication,
as suggested \cite{ln2}. Net current $I_{net}=(|I^+| - |I^-|)$ of
one such 30nm sample measured with gold sputtered contact at 28K
temperature is shown in inset of Fig. 2(b). This quenching
procedure reduces the mobility of dopants and contribution of
ionic conductivity in total current become much less even above
200K temperature.

Below 200K electronic nature of transport dominates and one
expects to observe $ ln \rho \propto (T_0/T)^{\beta}$ dependence
of resistivity. The exponent $\beta$ takes the value of $1/4$ for
Mott 3D VRH \cite{Mott} but approaches the value of $1/2$ if
Coulomb interactions create a gap as argued by Efros and
Shklovskii \cite {Efros}. Based on a heuristic calculation a cross
over function has been proposed as \cite{Amnon}

\begin{equation}\label{amnon}
    f(x)=\frac{1+[(1+x)^{1/2}-1]/x}{[(1+x)^{1/2}-1]^{1/2}}
\end{equation}
with $x=T/T_x$. This function exibit smooth crossover from Mott
$f(x)\propto x^{-1/4}$ for $x\gg 1$ to Efros-Shklovskii
$f(x)\propto x^{-1/2}$ for $x\ll 1$ behaviors. In Fig. 3(a) we
have shown variation of resistance of a quenched 30nm sample with
gold sputter contact as a function of temperature and a fit to Eq.
(1) that gives the value of $T_x$ as 9K. The deviation of data at
higher temperature from $f(x)$ is expected \cite{Amnon} due to
direct thermal activation. This fit clearly indicates that for the
measured temperature range (25K to 300K) we are in Mott's 3D
variable range regime as confirmed in the fit shown in the inset
of Fig. 3(a). It has been shown \cite{Int} that Mott's $T^{-1/4}$
VRH conduction law persists even in intrastate interacting
electron system. It may be noted that $T_0$ obtained in both
fittings is consistent \cite{Amnon}. We also observed $\beta=1/4$
in all the non-quenched samples with silver past contact and some
of the data and corresponding fits are shown in Fig. 3(b). For
30nm unquenched sample voltage goes into nonlinear region as the
resistance grows substantially at lower temperature ($<175K$).
Taking value of localization length ($\alpha^{-1}$) $\sim 0.2nm$
\cite{12,singh} the parameters density of states ($N(E_F)$),
hopping range ($R_{hop}$) and hop activation energy ($ W$) comes
out to be as $6.2\times 10^{22},7.68\times 10^{21},2.55\times
10^{21},1.24\times 10^{20}\; cm^{-3}eV^{-1}$ ; $0.45, 0.77, 1.02,
2.17\;nm$ and $40, 68, 88, 188\;meV$ for bulk, 200nm, 50nm and
30nm sample respectively.

The observed electrical transport properties of polypyrrole
nanowires and ratchet effect reported here can be explained by
considering a simple model having unconnected linear array of
nanospheres, representing the extent of the localized states, with
decreasing diameter ($d$) as $c$ reduces along the length of the
nanowires from $x=0$ to $x=L$ (refer Fig. 4(a)).  This variation
gives rise to gradual increase in the bare energy level spacing
and charging energy ($Q^2/{4\pi\epsilon_0\epsilon d}$) ($\epsilon$
is the static dielectric constant) with reducing value of diameter
($d$) of the nanospheres that hinder the symmetry of the VRH
charge transport mechanism \cite {14,15,16,17,18}. In Fig. 4(b) we
have quantified the inherent asymmetric hopping process in our
nanowires by considering forward and reverse bias hopping of
charge between two adjacent nanospheres of two different sizes
signifying different extent of localized sites. As a bias is
applied to the system, the charges start to hop from one site to
another. In a nanosphere the energy of an added electron depends
on both charging energy, and the bare energy level spacing \cite
{19}. The size asymmetry of adjacent nanosphere introduces
observed ratchet effect \cite {20,21,22} in the flow of charges
here. Let the energy of an added electron for the larger left and
smaller right nanospheres are $U_{cL}$ and $U_{cR}$ respectively
($U_{cL}<U_{cR}$). The effective barrier height on the right side
for hopping towards left is $\Delta_R$ and that for the left side
is $\Delta_L$ where $\Delta_R=\Delta_L- U_d$. The difference
charging energy $U_d=(U_{cR}-U_{cL})\equiv Q^2/(4
\pi\epsilon)\left(\frac{1}{d_R}-\frac{1}{d_L}\right)$ where $d_R$
and $d_L$ are diameter of right and left sphere respectively.
Under the application of positive (negative) bias the hopping
rates (transition probabilities) $\Gamma_R^\pm$ and $\Gamma_L^\pm$
can be expressed as
\begin{equation}\label{eqn-1}
\Gamma_R^\pm=\Gamma_0\;exp\{-(\Delta_R\mp 0.5 V
+\left|{\Delta_R\mp 0.5 V}\right|)/(2kT)\}
\end{equation}
from right to left and
\begin{equation}\label{eqn-2}
\Gamma_L^\pm= \Gamma_0\;exp\{-(\Delta_L\pm 0.5 V
+\left|{\Delta_L\pm 0.5 V}\right|)/(2kT)\}
\end{equation}
from left to right, where $V$ is the total potential drop across
the two sites \cite {18,23}. Current for positive (negative) bias
can be written as $I^\pm=Qb\left(\Gamma_R^\pm-\Gamma_L^\pm\right)$
where $Q$ is the amount of charge that hops and $b$ is the
distance between the two hoping sites. A very small value of $U_d$
can introduce significant asymmetry in the current.

A parameter $\phi$ defined as $\phi=(|I^+|-|I^-|)/(|I^+|+|I^-|)$
has been used to quantify the observed asymmetry where $|I^+|$ and
$|I^-|$ represent magnitude of forward and reverse current for
same magnitude of voltage, respectively. The measured values of
$\phi$ was found to decrease appreciably with increasing diameter
of nanowires - for example, at room temperature $\phi$ come out to
be around 0.3, 0.12, 0.012 and 0.008 for samples having nanowires
of nominal diameter of 30, 50, 100 and 200 nm respectively. The
values of $\phi$ become zero for bulk sample measured using same
geometry and electrical contacts. The values of $\phi$ was found
to increase with increasing temperature for all nanowires, as
shown in the lower inset of Fig. 4(c) for the samples having
nanowires of 30 nm diameter. In our model even for $U_d=0.4$ meV
we get $\phi= 0.2$ at 300K (taking barrier height$=188meV$,
voltage drop across a barrier $=2meV$). The observed increase in
$\phi$ with increasing temperature (Fig. 4(c) lower inset) is
generally not expected in the hopping process. In the first
approximation $\phi$ should have been independent of temperature
as in polypyrrole dielectric constant $\epsilon$ is known to be
proportional to $T^{-1}$ \cite{singhrp} and that would have made
$U_d$ proportional to $T$. In Fig. 4(c) we have shown variation of
$\epsilon$ of a 30nm sample with temperature obtained from the
peak in the capacitance value measured as a function of frequency
\cite{singhrp} (refer upper inset in Fig. 4(c)). We get $\epsilon
\propto T^{-1.4\pm 0.06}$ and this in turn provide a qualitative
explanation of increase in $\phi$ with temperature. For example in
the above calculation $U_d$ now becomes 0.057meV at 74K and that
gives a value of $\phi$ as 0.02. This is close to the experimental
value of 0.01 shown in lower inset of Fig. 4(c). It should be
mentioned here that $\epsilon$ of water solvent was found to be
proportional to $T^{-1.5}$ instead of $T^{-1}$ as dipole rotation
get hindered \cite{chaiken} and similar hinderance may be existing
for our polypyrrole nanowires formed in confined geometry.

In conclusion, the observed ratchet effect in electrical transport
of polypyrrole nanowire could be explained with a simple model
having growing extent of localization sites along the nanowire.
One should be able to make charging rectifier in polymer nanowires
simply by reducing the radius of the nanospheres further from the
present value of $30$ nm used here.

Authors would like to thank Prof. A. K. Sood for valuable
discussions. Authors are grateful to Prof. P. Chakraborty and
Prof. A. Ghosh for their help in SIMS and capacitance
measurements, respectively.

\begin{figure}

{\bf Figure captions:} \caption{(a) Schematic diagram of the
set-up used to prepare nanowires of conducting polymer in
nanopores of membranes. (b) SEM image of a bundle of nanowire
formed in this process. (c) $Cl$ and $N$ profiles obtained from
SIMS measurements for 30nm and 200nm nanowires are shown in the
upper panel as a function of length. Ratio between these two
profiles normalized to maximum doping of 0.33 is shown in the
lower panel.}

\caption{(a) Difference in resistance value for forward and
reverse bias current of 30 nm nanowires shown before and after
switching the polarity of the bias at t=50sec. Electrical
connections for measuring current-voltage characteristics have
been shown in the inset. (b) Voltage developed across the
nanowires as a function of forward (dashed line) and reverse
(continuous line) bias current taken at various temperatures. Low
temperature data has been multiplied by constant factors, as
shown, to use the same scale. Higher temperature data have been
shifted vertically for clarity. Net current is plotted as a
function of voltage for liquid nitrogen quenched 30nm sample
having gold contact (inset).}

\caption{(a)Resistance vs temperature plot and fit to Eq. (1) to
determine crossover from $T^{-1/4}$ regime to $T^{-1/2}$ regime.
Same data plotted in inset to show that $T^{-1/4}$ law is valid in
our measurement range. (b) Same plot as inset of (a) for samples
having different diameters.}

\caption{(a) Schematic representation of expected increase of the
extent of localized states along the length ($L$) of nanowires
represented by unconnected conducting nanospheres of increasing
diameters is shown. Difference in charging energy due to an added
electron on these nanospheres has been indicated. (b) Hopping
model used to explain the resistance asymmetry is shown. (c)
Dielectric constant obtained from frequency dependent capacitance
data (shown in upper inset) as a function of temperature is
plotted in Log-Log scale. The parameter $\phi$ that quantifies the
ratchet effect is plotted as a function of temperature (in lower
inset).}

\end{figure}

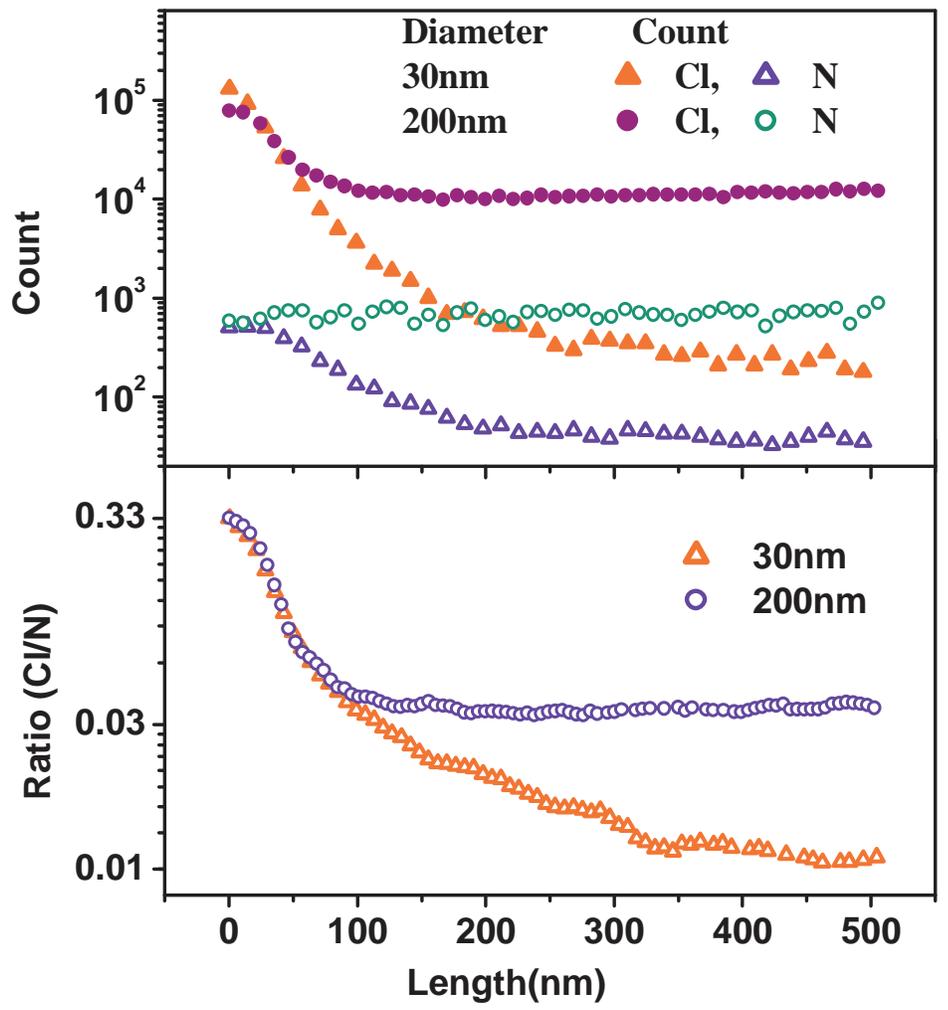
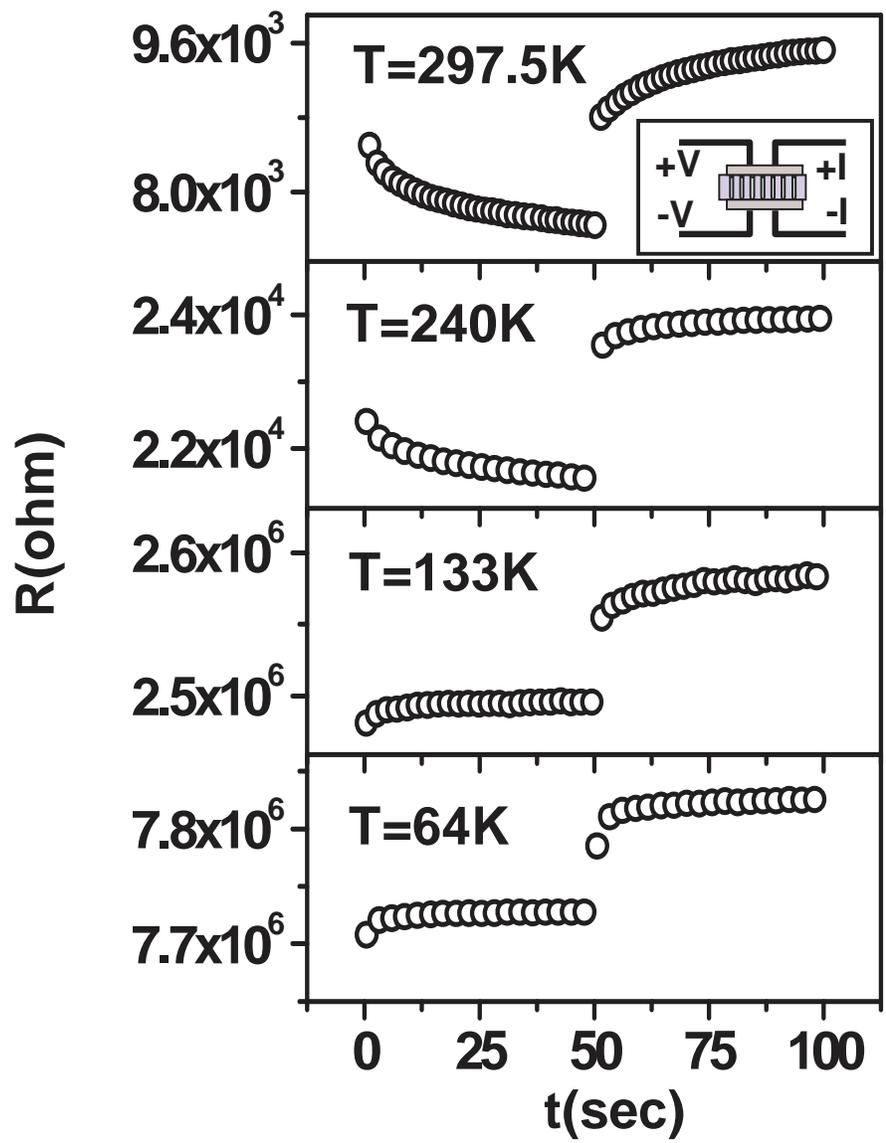
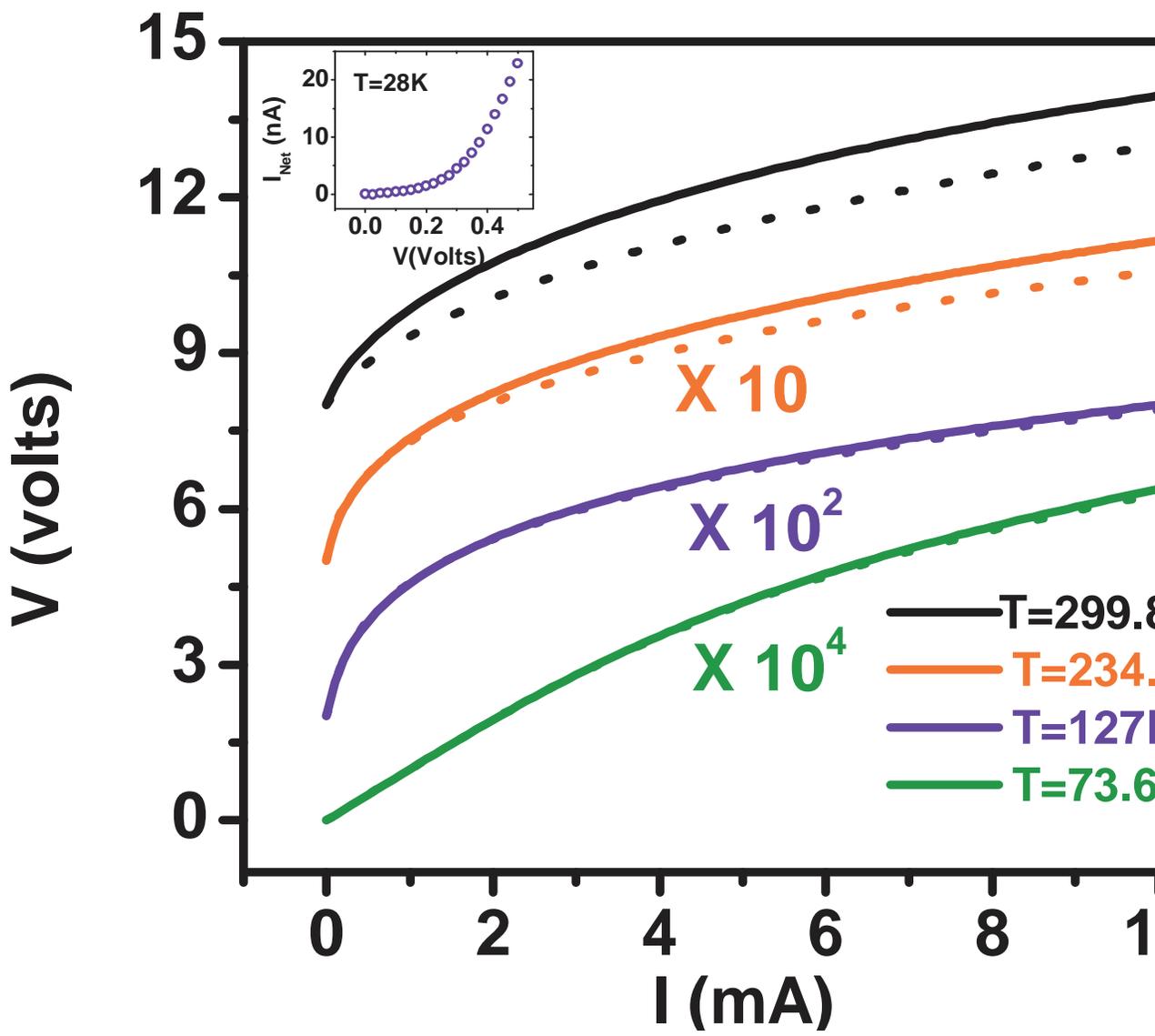
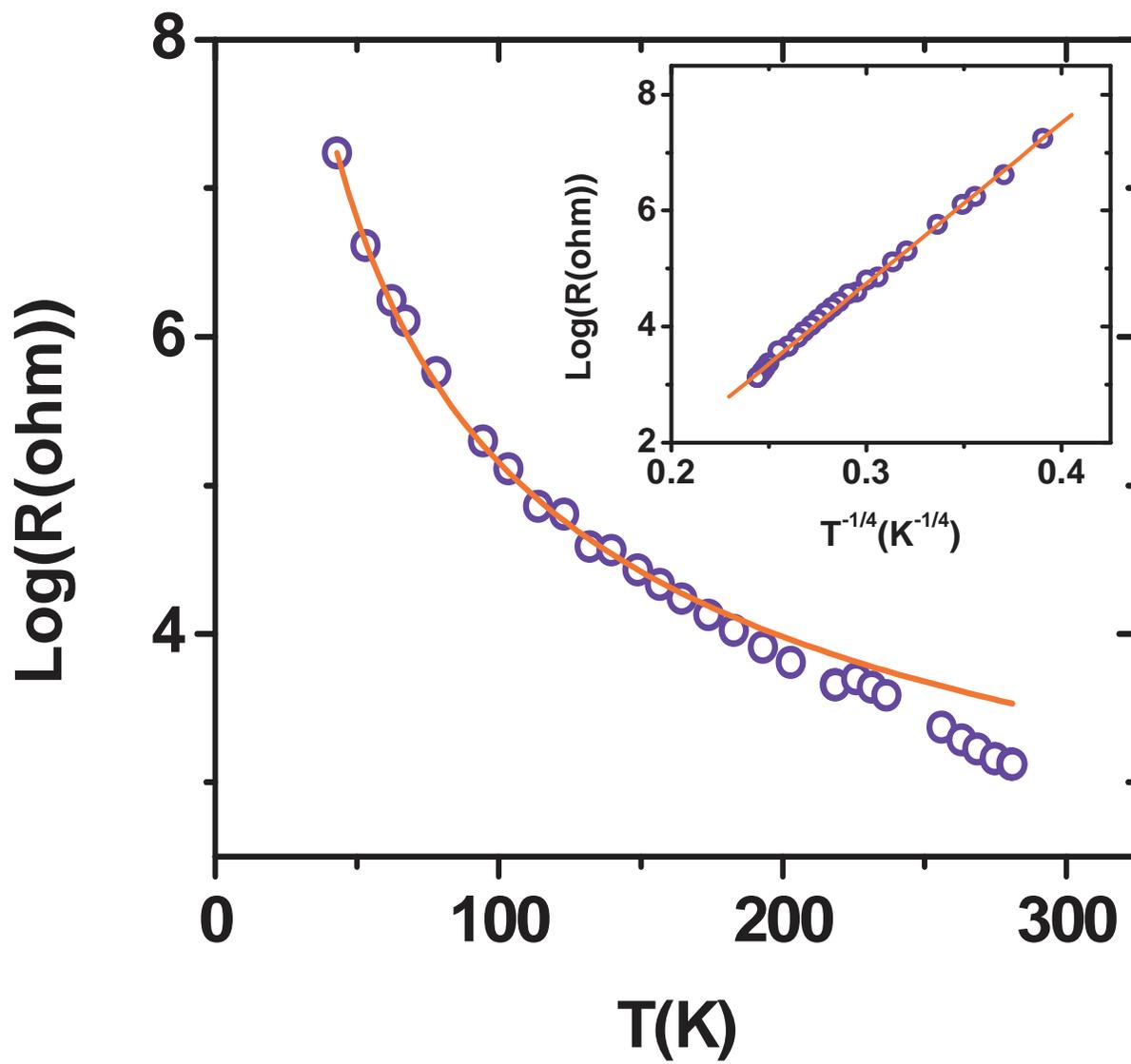
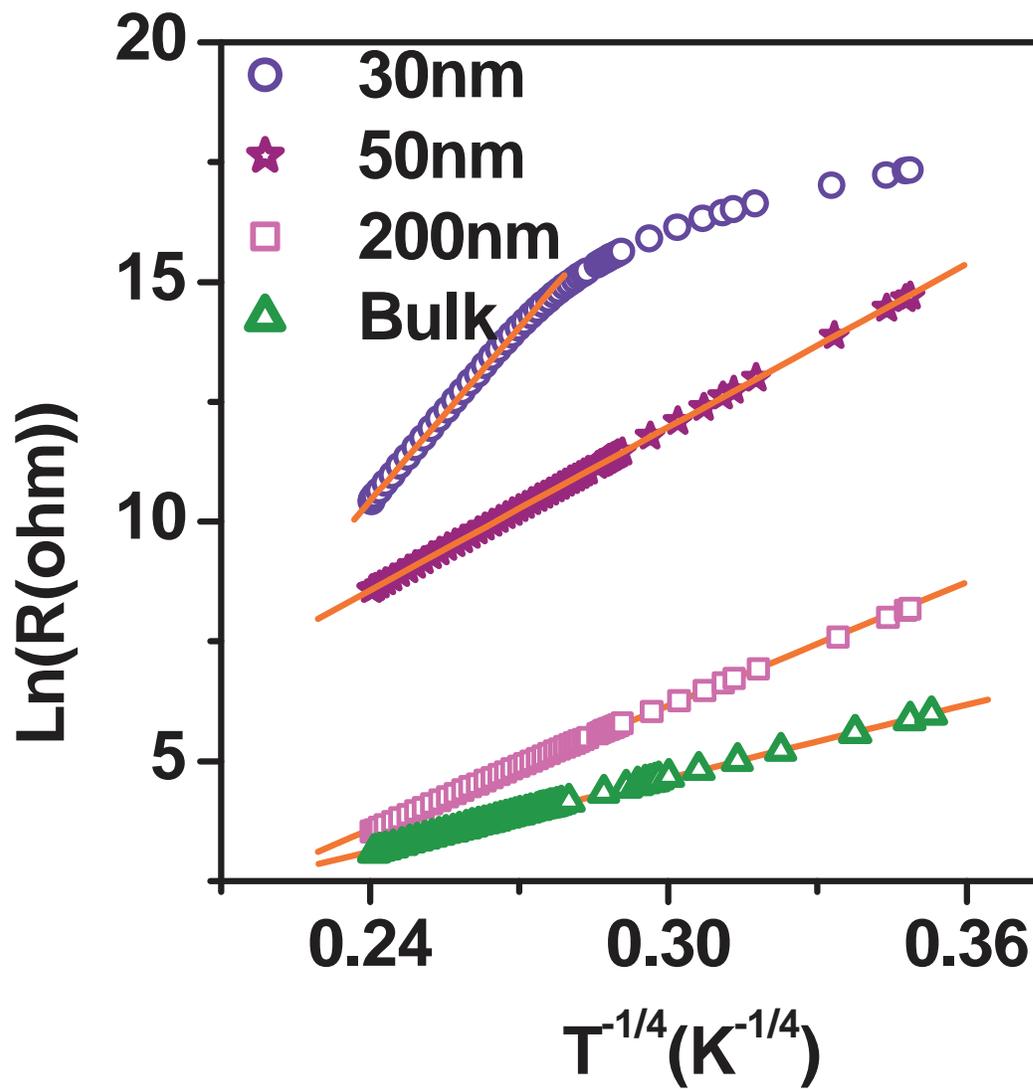
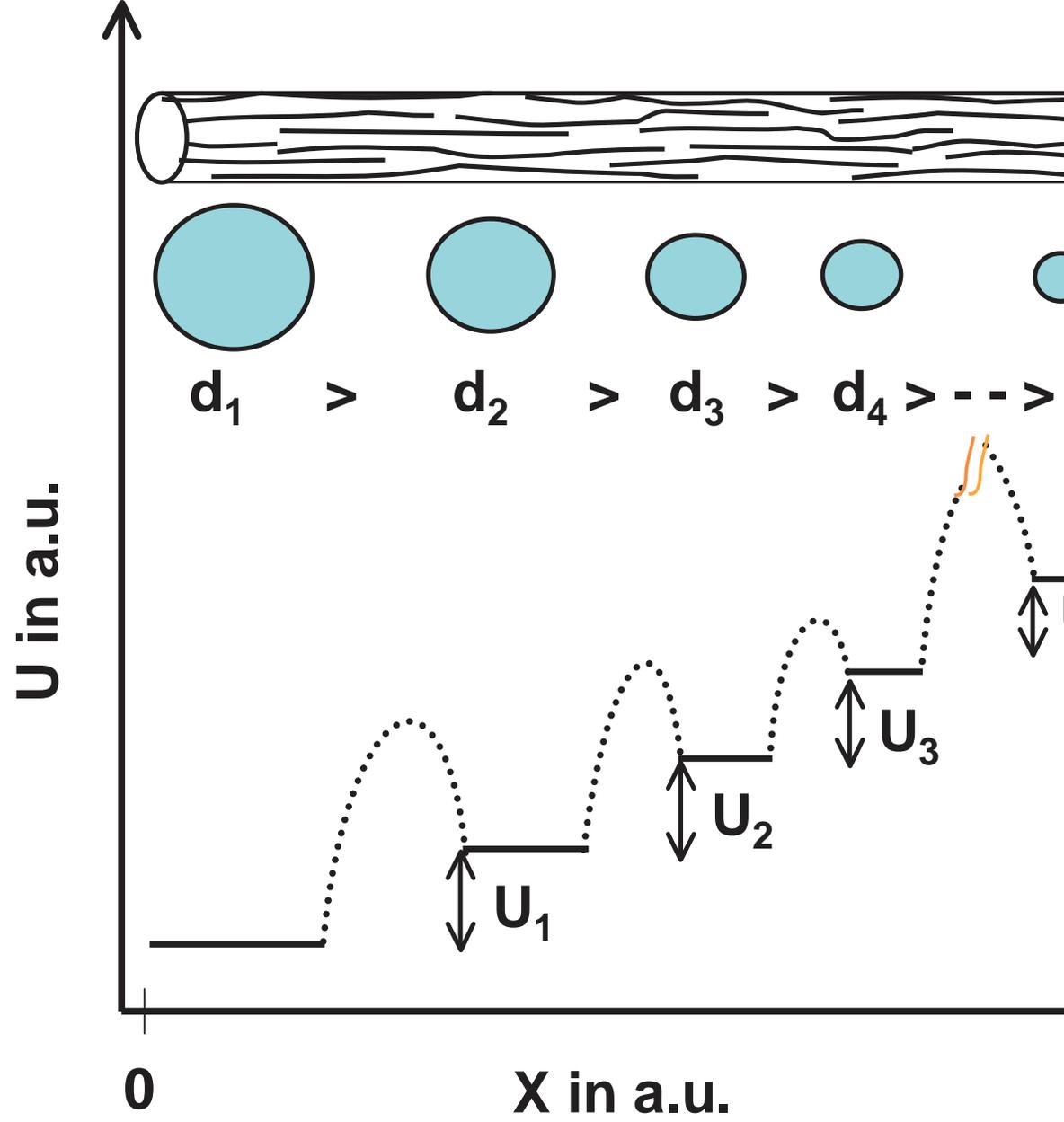
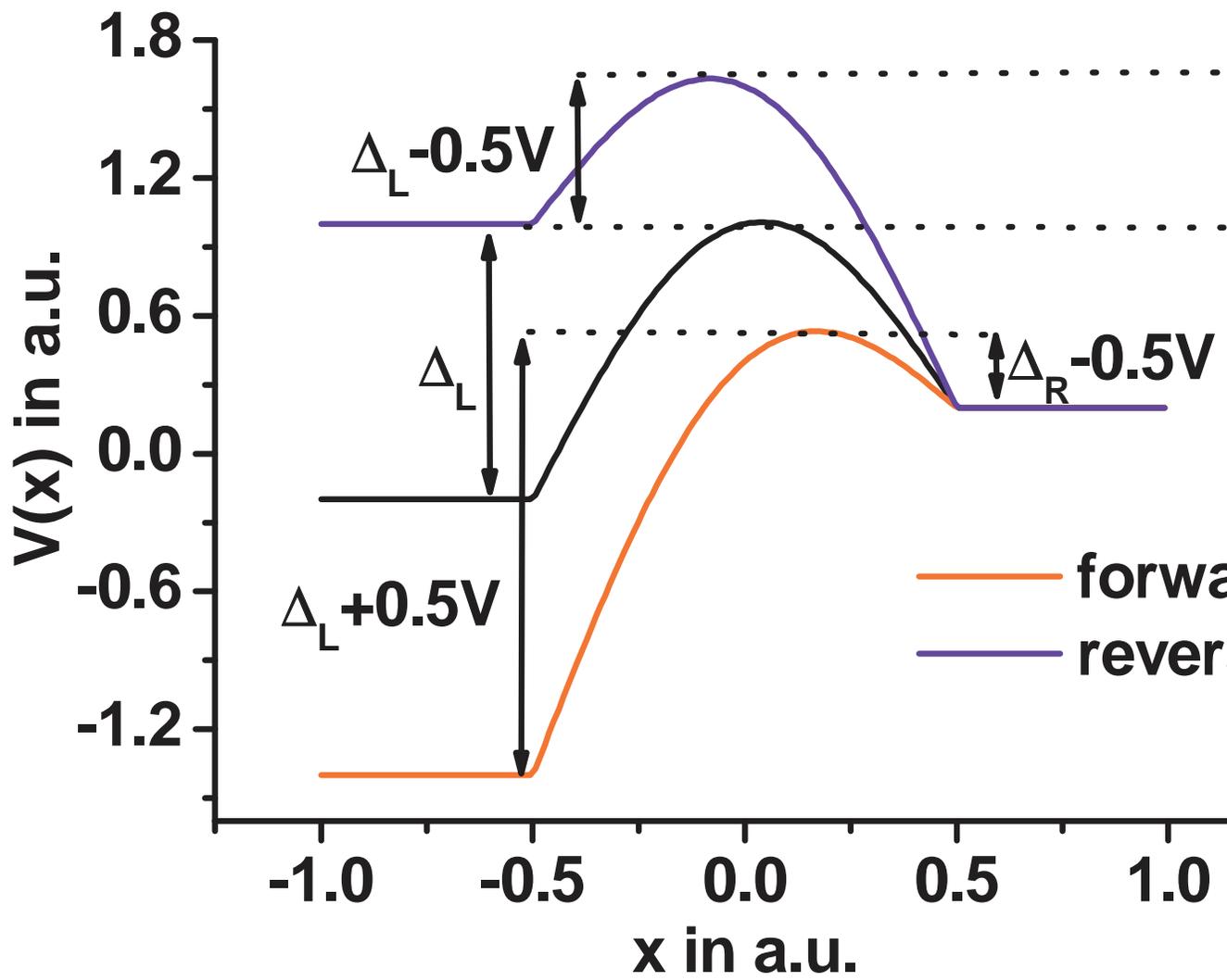
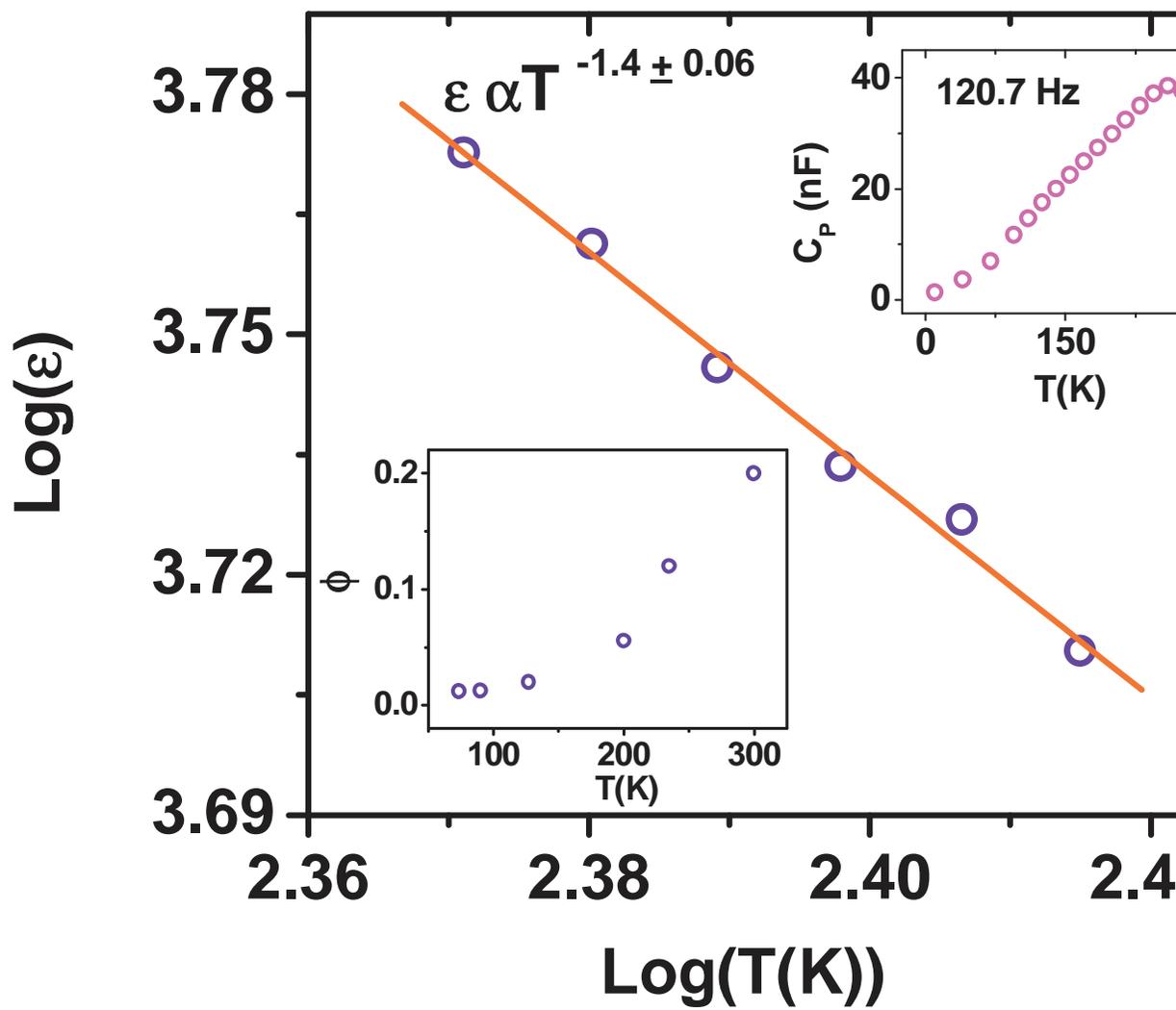
\end{document}